\def\a{a}
\def\A{A}
\def\b{\overline \a}
\def\B{\overline \A}
\def\L{\ln\Lambda}

\def\E{{\cal E}}
\def\K{{\cal K}}
\def\P{{\cal P}}
\def\H{{\cal H}}
\def\e{{\epsilon}}
\def\k{{\kappa}}
\def\lOA{{1OA}}
\documentstyle[sprocl,graphicx]{article}

\def\be{\begin{equation}}
\def\ee{\end{equation}}
\def\bea{\begin{eqnarray}}
\def\eea{\end{eqnarray}}
\begin{document}

\title{COMPETITION OF FERROMAGNETIC AND ANTIFERROMAGNETIC ORDER
IN THE SPIN-1/2 XXZ CHAIN AT FINITE TEMPERATURE}

\author{K. Fabricius}

\address{Physics Department, University of Wuppertal,\\
42097 Wuppertal, Germany}

\author{A. Kl\"umper}

\address{Institut f\"ur Theoretische Physik,
Universit\"at zu K\"oln, Z\"ulpicher Str. 77,\\ 
50937 K\"oln 41, Germany}

\author{B. M. McCoy}

\address{Institute for Theoretical Physics, State University of New York,
Stony Brook,\\NY 11794-3840}


\maketitle\abstracts{An analytical study is presented of the crossover
in the gapless attractive $XXZ$ chain
from antiferromagnetic to ferromagnetic behaviour at low to high
temperature, respectively. In particular, an analytic formula
for the crossover in the long distance asymptotics and explicit results 
for the nearest-neighbour longitudinal correlation are obtained.
We also provide results for the specific heat and magnetic susceptibility
for various anisotropies.
}
  
\section{Introduction}\label{subsec:intro}

In this paper we pursue the question to which extent the partially
anisotropic Heisenberg chain with nearest-neighbour coupling
\be
H_{XXZ} = {1 \over 2} \sum_{j=1}^L {(\sigma_j^X \sigma_{j+1}^X
+\sigma_j^Y \sigma_{j+1}^Y+\Delta \sigma_j^Z \sigma_{j+1}^Z)}.
\label{Hamilt}
\ee
is described by conformal field theory \cite{car,affb}.
It is well known that (\ref{Hamilt}) for anisotropy $-1<\Delta<+1$ 
has a critical groundstate \cite{yya,yyb,yyc,lp,fog}, 
however with quite different physics 
regarding the elementary excitations \cite{jkm,klz}. There are gapless
elementary spin excitations
throughout $-1<\Delta<+1$, but additional bound states
in the ``attractive'' parameter regime $-1<\Delta<0$. Note that $\Delta=-1$
corresponds to the isotropic ferromagnetic Heisenberg chain with fully
polarized groundstate.

Correlation functions are difficult 
to calculate, analytical results are known for the nearest-neighbour 
longitudinal correlation \cite{jkm}
$\langle\sigma_j^Z \sigma_{j+1}^Z\rangle_{T=0}$ (which will be
generalized to finite $T$ in Sec.~\ref{sec:crossover}) as well as
field-theoretical results for the long-range behaviour of correlations
for $T=0$ \cite{lp,fog} and small non-zero $T$ \cite{BogK89,kbi}. 
Amazingly, the nearest-neighbour correlation
$\langle\sigma_j^Z \sigma_{j+1}^Z\rangle_{T=0}$ is always negative
for all anisotropies $-1<\Delta<+1$. In addition, the longitudinal 
correlation $\langle\sigma_0^Z \sigma_{r}^Z\rangle_{T=0}$
for large distances $r$ is negative in the attractive regime. 

The finite temperature results for 
$\langle\sigma_0^Z \sigma_{r}^Z\rangle$ 
as found in numerical studies are quite rich \cite{FabMc}. For fixed
separation $r\not= 0$ a transition from negative
to positive values appears for some temperature $T_0(\Delta,r)$.
In \cite{FabMc} the question was raised whether $T_0(\Delta,r)$
possesses a well defined (and non-zero) value in the limit $r\to\infty$.
Alternatively, in an asymptotic expansion in exponentials
\be
\langle\sigma_0^Z \sigma_{r}^Z\rangle=
A \exp\left(-r/\xi\right)+...,
\ee
the temperature dependence of the amplitude $A$ of the dominant term
was addressed and it was argued that there is a sign change for some 
well defined temperature $T_0(\Delta)$ ($=T_0(\Delta,\infty)$). 
Nothing like this is observed
in the ``repulsive'' regime where correlations show simple antiferromagnetic
oscillations. 

In the current work we want to understand analytically the sign change
phenomenon.
In particular, we aim at an analytic formula for $T_0$, see (\ref{T0}). 
For two reasons, these questions are not purely academic.
First, the ``attractive'' $XXZ$
chain is often considered as a system with ferromagnetic interactions,
however with antiferromagnetic groundstate, i.e. vanishing magnetization. 
Our analysis is aimed 
at a resolution of the somewhat unintuitive picture in the way of a crossover
from dominant ferromagnetic correlations at high to dominant antiferromagnetic
correlations at low temperatures. Second, we address the fundamental
problem of the additional
energy scale that has to enter the description of the 
``attractive'' regime, but is absent in the ``repulsive'' case.
Naturally, we will be led to pay attention to the existence of bound states.

In Sec.~\ref{subsec:groundstate} we review the known groundstate
properties of the $XXZ$ chain, and in Sec.~\ref{subsec:QTM}
we set up our formalism for finite temperatures on the basis of
the quantum transfer matrix. In Sec.~\ref{sec:crossover}
we address the problem of crossover phenomena in correlation
functions and macroscopic properties. Finally, in Sec.~\ref{sec:Concl}
we summarize our results and list the open problems.

\section{Groundstate Properties}\label{subsec:groundstate}
In (\ref{Hamilt}) the anisotropy $\Delta$ is conveniently parametrized by 
\be
\Delta=\cos\gamma
\ee
where the range $0\le\gamma\le\pi/2$ corresponds to repulsive interactions
and $\pi/2\le\gamma<\pi$ to attractive interactions. The elementary 
excitations on the (antiferromagnetic) groundstate are free states 
with dispersion relation
\be
\epsilon_f(k)=v \sin k,\qquad 0\le k\le\pi,
\label{spinons}
\ee
and ``sound'' velocity
\be
v={\sin\gamma\over\gamma} \pi.
\ee
In addition to these ``free states'' there occur
``bound states'' in the
attractive regime $\pi/2\le\gamma<\pi$. Depending on the anisotropy
there are different bound states labeled by an integer $\mu$ in the range
$1\le\mu\le\left[{\gamma\over\pi-\gamma}\right]$ with dispersion
\cite{jkm,klz}
\be
\epsilon_b(k)=
v\, 2 \sin{k\over 2}\sqrt{1+a_\mu^2\sin^2{k\over 2}},
\qquad a_\mu=\cot\left(\mu{\pi-\gamma\over\gamma}{\pi\over2}\right).
\label{boundspinons}
\ee
and $0\le k\le\pi$. Note the same velocity of the bound state dispersion
as for the free state dispersion. However, for finite momentum the
energy of the bound states is {\em larger} than that of the free states.

%Add
%compactified boson
For the field theoretical description of the $XXZ$ chain the occurence 
of bound states in the attractive regime does not have any 
fundamental consequences, because of identical velocities of free and bound 
states in the long-wave limit. Within the bosonization 
approach the $XXZ$ chain is mapped to a
Sine-Gordon model where the interaction term is argued to be infrared 
irrelevant (resulting in a Gaussian model). In this way correlation 
functions for $T=0$ and small $T>0$ with asymptotic behaviour
\be
C_r\sim C \cos(P_0 r)
\left({{\pi\over \beta v}\over\sinh{\pi\over \beta v}r}\right)^{2x},
\label{asCFT}
\ee
are derived \cite{BogK89,kbi}. 
To our knowledge the breakdown of this picture at higher
temperatures in general, and the occurence of a new temperature scale
in the attractive regime (with bound states)
in particular has not been studied for the Sine-Gordon model.

The scaling dimension $x$ and ``lattice momentum'' $P_0$
are generally different for
different correlation functions and given by a formula
obtained from lattice calculations 
\be
x={1-\gamma/\pi\over 2}S^2+{1\over 2(1-\gamma/\pi)}m^2+
k,\qquad P_0=(S-m)\pi,\label{scaldim}
\ee
where $S$, $m$, $k$ are integers corresponding to spin, lattice momentum
and position in the conformal tower. For the longitudinal spin-spin
correlation function the selection rules enforce $S=0$, but leave
open the values for $m$ and $k$. A simple inspection shows that
$(m,k)=(1,0)$ ($=(0,1)$) gives the smallest scaling dimension $x$
for the repulsive (attractive) regime. 

\section{Finite Temperatures}
\label{subsec:QTM}
\subsection{Quantum Transfer Matrix}
For a treatment of finite temperatures we employ a convenient 
transfer matrix approach \cite{Suz85,SuIn87,Suz90,klu}. 
To this end the quantum chain at finite
temperature is mapped via a Trotter-Suzuki decomposition
onto a two-dimensional classical 
model on a square lattice of width $L$ (= chain length) and height
$N$ (= Trotter number) and staggered interactions. The free energy and
the decay of static correlation functions is completely described 
by the largest eigenvalue ($\L_0$) and the next-largest eigenvalues 
($\L_i$) of the 
column-to-column transfer matrix
\bea
f&=&-{1\over\beta}\lim\L_0\nonumber\\
C_r&=&A_1 \left({\Lambda_1\over\Lambda_0}\right)^r
+A_2 \left({\Lambda_2\over\Lambda_0}\right)^r
+... .
\label{expansion}
\eea
In the limit $N\to\infty$ this matrix
is referred to as the quantum transfer matrix (QTM) being the closest
analogue to the usual transfer matrix of classical spin chains.

For the $XXZ$ chain integrability is manifest at the level of the QTM 
in the following way. There is a commuting family of matrices $T(x)$
generated by the spectral parameter $x$. The QTM is identical
to $T(0)$. As most of the physical properties are directly given
by the logarithm of $T(0)$ we define $\H=-\ln T(0)$. The lowest eigenvalue
of $\H$ gives $\beta f$ where $f$ is the free energy per chain site.
Furthermore, the QTM enjoys translational invariance 
along the vertical axis within the two-dimensional geometry. The 
corresponding momentum $\P$ operator is generated by $T(x)$ through 
differentiation. In summary we have
\bea
\H&=&-\ln T(0),\nonumber\\
\P&=&-i\sin\gamma{d\over dx}\ln T(x)|_{x=0}.
\eea
Note the similarity to the integrability 
structure of the Hamiltonian $H$ and momentum operator $P$
with one important difference. The momentum operator $P$
is given by the row-to-row transfer matrix 
and $H$ is given by its derivative. Consequentially, we will often
observe a formal correspondence \hbox{$\H$, $\P$} $\longleftrightarrow$
\hbox{$i P$, $i H$}. Of course, in a suitable continuum limit this is
easily understood as Hamiltonian and momentum operator are evolution 
operators with respect to orthogonal directions.

Next, we review the already known eigenvalue equations for the
operator $T(x)$. The corresponding eigenvalues
are denoted by $\Lambda(x)$ from which the eigenvalues $\E$ and $\K$ for
$\H$ and $\P$ are directly obtained 
\bea
\E&=&-\L(0),\nonumber\\
\K&=&-i\sin\gamma{d\over dx}\L(x)|_{x=0},
\label{eigenvaluesEK}
\eea
the latter one taking all integer multiples of $2\pi/\beta$. This 
discretization of $\K$ naturally entails a discrete spectrum for $\E$.

\subsection{Eigenvalues and Non-Linear Integral Equations}
The largest eigenvalue is given by \cite{klu}
\be
\L(x)=-\beta e_0(x)+{1\over 2\gamma}
\int_{-\infty}^\infty{\ln[\A\B(y)]\over\cosh{\pi \over \gamma}(x-y)}dy.
\label{eigenxxz}
\ee
where $e_0$ is the groundstate energy and $\A=1+\a$, $\B=1+\b$
are solutions to the following set of non-linear integral equations (NLIE)
\bea
\ln \a(x) 
&=& -{\beta v\over\cosh{\pi \over \gamma}x}
+{\pi\over2(\pi-\gamma)}{\beta h} \nonumber\\
&&+\int_{-\infty}^\infty\left[k(x-y)\ln \A(y)
-k(x-y-i\gamma+i\epsilon)\ln\B(y)\right]dy.
\label{NLIE}
\eea
$h$ is the externally applied magnetic field and the integration kernel 
is given by
\be
k(x)={1\over 2\pi}
\int_{-\infty}^{\infty}{\sinh\left({\pi\over 2}-\gamma\right)k\cos(kx)
\over2\cosh{\gamma\over 2}k\sinh{\pi-\gamma\over 2}k}dk.
\ee
The corresponding equation for $\b$ is obtained from 
Eq.~(\ref{NLIE}) 
by exchanging $i \to -i$, $h \to -h$ and $\a,\A \leftrightarrow \b,\B$.
These equations will be studied numerically in the next section yielding
results for the specific heat, magnetic susceptibilities, and
nearest-neighbour correlation for various
anisotropies $\Delta$ and wide temperature ranges. 

The next-largest eigenvalues describing the decay of longitudinal spin-spin
correlation functions, i.e. eigenvalues with spin quantum number
$S=0$, are given by \cite{klu,kwz}
\bea
\L(x)&=&-\beta e_0(x)+
\ln\left[\tanh{\pi\over 2\gamma}(x-\theta_1)
\tanh{\pi\over 2\gamma}(x-\theta_2)\right]\nonumber\\
&&+{1\over 2\gamma}
\int_{\infty}^\infty{\ln[\A\B(y)]dy\over\cosh{\pi \over \gamma}(x-y)},
\label{eigexc}
\eea
where $\A$ and $\B$ are determined from
\bea
\ln \a(x)&=& -{\beta v\over\cosh{\pi \over \gamma}x}
+\pi i+{\pi\over 2(\pi-\gamma)}\beta h 
+\ln{\sinh{\pi\over\pi-\gamma}(x-(y_0+i\gamma/2))
\over\sinh{\pi\over\pi-\gamma}(x-(y_0-i\gamma/2))}
\nonumber\\
&&-K(x-(\theta_1+i\gamma/2))-K(x-(\theta_2+i\gamma/2))\nonumber\\
&&+\int_{\infty}^\infty\left[k(x-y)\ln \A(y)
-k(x-y-i\gamma+i\epsilon)\ln\B(y)\right]{\rm d}y.
\label{excited}
\eea
and $K(x)$ is defined by $K(x)'=2\pi i k(x)$, or explicitly
\be
K(x)=
i\int_{-\infty}^{\infty}{\sinh\left({\pi\over 2}-\gamma\right)k\sin(kx)\over
2k\cosh{\gamma\over 2}k\sinh{\pi-\gamma\over 2}k}dk.
\ee
There appear three parameters $\theta_1$, $\theta_2$, and $y_0$ in the upper
equations corresponding to ``hole positions'' (both $\theta_{1,2}$
on the real axis, or forming a complex conjugate pair)
and one ``complex rapidity'' (with Im$(y_0)=\pi/2$) of the underlying
Bethe ansatz pattern for the QTM. There are higher-lying states 
satisfying similar integral equations, however with more parameters 
$\theta_i$ and $y_j$.

The parameters $\theta_1$, $\theta_2$, and $y_0$
do not take arbitrary (continuous) 
values, they have to satisfy the coupled equations 
\be
\a(\theta_1+i\gamma/2)=
\a(\theta_2+i\gamma/2)=\a(y_0+i\gamma/2)=-1,
\label{subsidiary}
\ee
leading to a quantization of the eigenvalues.
In general the equations (\ref{excited},\ref{subsidiary})
have to be solved numerically in order to deal with the problem
of non-linearity. For the case of the largest and next-largest 
eigenvalues an iterative approach proved useful
showing convergence within a numerical accuracy of $10^{-6}$
already after approx.~10 steps, see ref. \cite{klu} for the non-critical
cases with $\Delta>1$ and $\Delta<-1$, and ref. \cite{klunew} for the critical
case $\Delta=1$.
A comparable numerical analysis for $-1<\Delta<1$ is in progress, 
however not yet completed. For this reason
we apply an analytical study within a reasonable approximation to 
(\ref{excited},\ref{subsidiary}).

\subsection{Conformal Field Limit for Low Temperatures}
In \cite{klu} the equations (\ref{eigexc},\ref{excited}) were treated in the
low-temperature limit. Despite the apparent complexity
of the non-linear integral equations, quite a universal picture
evolved. A certain symmetry of the integration kernel allowed
for analytic manipulations avoiding the necessity of an explicit 
solution of the non-linear equations. Each eigenvalue could be 
written in terms of dilogarithmic
integrals which resulted in the explicit formula
\be
\L=-\beta e_0-{2\pi\over v} T (x-c/12)+o(T^2)+i P_0,\label{result}
\ee
where the central charge is $c=1$ and the scaling dimension $x$ 
and lattice momentum $P_0$ are given by (\ref{scaldim}).
From (\ref{expansion},\ref{result}) we see
\be
C_r\sim \cos (P_0r) \e^{\hbox{$ -{2 \pi\over v}x T r$}}.\label{asympCorr}
\ee
which coincides with the conformal field theory result (\ref{asCFT}) in
the low-temperature limit. 

On one hand we see that (\ref{eigexc},\ref{excited}) recover 
CFT, on the other hand, the non-linear integral equations are 
not restricted to low temperatures, but go beyond. It is this
behaviour that will be studied in the next sections.

\subsection{Solution in Lowest Order}\label{subsec:first}
A principal treatment of the NLIE is based on iterations. For instance in the 
case of (\ref{NLIE}) a reasonable 
initial choice for the functions $\a$ and $\b$ is
just $\a_0=\b_0=0$. Inserting this into the right hand side of 
(\ref{NLIE}) leads to 
\be
\ln \a_1(x) 
= -{\beta v\over\cosh{\pi \over \gamma}x}.
\label{auxfree}
\ee
This process should be continued {\it ad infinitum}, however we
content ourselves with the approximation $\a_1$ for the function $\a$.
Inserting $\a_1$ and $\b_1$ into (\ref{eigenxxz}) we obtain the
{\it first order approximation} (\lOA) in the sense of an iterative procedure
to the largest eigenvalue of the QTM which already reproduces
correctly: (i) the low-temperature asymptotics 
$\L=-\beta e_0+c({\pi T/ 6v})$ with central charge $c=1$, {\it and}
(ii) the high-temperature behaviour $\L=\ln 2$. This success
may be understood in the following way. At low temperatures the
corrections described by the integral terms in (\ref{NLIE}) are small,
and at high temperatures both of them cancel each other.
We conclude that the (\lOA) is a respectable approximation useful for
the entire temperature range.

\subsubsection{Free States}
Next, we study the excitations (\ref{excited},\ref{subsidiary}) in (\lOA)
yielding
\be
\L(x)=
\ln\left[\tanh{\pi\over 2\gamma}(x-\theta_1)
\tanh{\pi\over 2\gamma}(x-\theta_2)\right],
\label{eigfree}
\ee
where we have dropped the common offset $-\beta e_0(x)$ in (\ref{eigexc}).
From this and (\ref{eigenvaluesEK}) we derive the eigenvalues
\bea
\E&=&\e_1+\e_2,\nonumber\\
\K&=&\k_f(\e_1)+\k_f(\e_2),\qquad \k_f(\e)=\pm v\sinh\e,
\eea
where the parameters $\theta_i$ have been parametrized by $\e_i (\ge 0)$
such that $\k_i$ is a unique function of $\e_i$ except for the sign in 
the $\k_f$ dispersion which is given by the sign of the parameter
$\theta_i$. 
Note the similarity of the ``momentum-energy'' dispersion $\k_f(\e)$ 
for the ``free states'' of the QTM with the dispersion (\ref{spinons})
\be
\k_f(\e)=-i\epsilon_f(i\e).
\ee
We know
already from general principles that $\K$ is strictly equal to an integer
multiple of $2\pi/\beta$. This does not necessarily
imply a similar property for each individual $\k_i=\k_f(\e_i)$, 
however values close to multiples of $2\pi/\beta$
are generally taken. 

We see this most easily for the case $\k_1=-\k_2$, i.e.~$\theta_1=-\theta_2$.
The quantization condition (\ref{subsidiary}) for $y_0$ requires $y_0=\infty$.
If we apply the (\lOA) treatment to (\ref{excited}),
the quantization condition (\ref{subsidiary}) simply reads
$\ln \a(\theta_i+i\gamma/2)=i{\beta v/\sinh{(\pi / \gamma)}\theta_i}=$
odd multiple of $\pi i$. Using this when inserting (\ref{eigfree})
into (\ref{eigenvaluesEK}) we directly find $\k_1=-k_2=2\pi/\beta$.

If both $\k_1$ and $\k_2$
take the smallest possible values of same sign and the 
$XXZ$ chain is in the attractive regime $\pi/2\le\gamma<\pi$
the quantization
conditions impose complex values for the parameters $\theta_{1,2}$.
This leads us to the study of bound states of the QTM.

\subsubsection{Bound States}
Here we content ourselves to a motivation of why in 
the attractive regime 
complex solutions to (\ref{subsidiary}) close to
\be
\theta_{1,2}=\theta\pm i\left(\gamma-{\pi\over 2}\right),
\label{boundcond}
\ee
appear. To this end
let us assume that $\theta_1$ has a positive imaginary 
part and we calculate $\ln\a(\theta_1+i\gamma/2)$ from (\ref{excited}).
For low temperatures the dominant term is 
$-\beta v/\cosh{(\pi / \gamma)(\theta_1+i\gamma/2)}$ whose real
part tends to $+\infty$ in the limit $\beta\to\infty$. The only term 
matching this divergence is the $y_0$ term in the first line of
(\ref{excited}) if $\theta_1-y_0=\gamma-\pi$. Likewise we conclude 
for $\theta_2$ with negative imaginary part the condition
$\theta_2-y_0+\gamma=0$. Using Im$y_0=\pi/2$ we write
$y_0=\theta+i\pi/2$ and arrive at (\ref{boundcond}). 

We like to note that for the case of higher-lying states involving more
than one complex rapidity $y_0$ there are additional bound states.
These always involve two complex conjugate parameters $\theta_{1,2}$
and a ``string'' of complex rapidities $y_j$ with minimum number 1 and
maximum number $[{\gamma/(\pi-\gamma)}]$, a situation similar
to that of the Hamiltonian. However, for the crossover phenomenon in
the correlation functions the case explicitly studied above
is sufficient.

The momentum-energy dispersion is obtained from 
(\ref{eigfree},\ref{boundcond}) inserted into 
(\ref{eigenvaluesEK}) and eliminating
$\theta$. More easily we may use the relation to (\ref{boundspinons})
$\k_b(\e)=-i\epsilon_b(i\e)$. With $\mu=1$ we find
\be
\kappa_b(\epsilon)=\pm
v\, 2\sinh{\e\over 2}\sqrt{1-a^2\sinh^2{\e\over 2}},
\qquad a=\cot\left({\pi-\gamma\over\gamma}{\pi\over2}\right).
\ee

\section{Crossover Phenomena}
\label{sec:crossover}
In this section we are going to investigate various crossover phenomena
in correlation functions as well as macroscopic properties
of the $XXZ$ chain in the attractive regime. Of prime 
interest is the calculation of the crossover temperature as function of
the anisotropy parameter $-1<\Delta\le0$.
\subsection{Crossover in Correlation Functions}
With the knowledge of Sec.~\ref{subsec:first}
we are in the position to examine the crossover of
the asymptotics of the longitudinal correlation function from 
ferromagnetic behaviour at high temperature to antiferromagnetic
behaviour at low temperature. In particular we want to derive an
analytic formula for the crossover temperature as function of the
anisotropy $\Delta$.
In principle such a crossover may occur under two separate circumstances
\begin{itemize}
\item{(i) the prefactor $A_1$ of the dominant term in (\ref{expansion}),
i.e. that with longest correlation length $\xi_1=1/(\E_1-\E_0)$,
turns zero for some temperature.}
\item(ii) the leading and next-leading terms in (\ref{expansion}) have 
prefactors $A_1$ and $A_2$ of different sign, and a crossover of
the correlation lengths $\xi_1$ and $\xi_2$ occurs for some
temperature.
\end{itemize}
Unfortunately, we cannot calculate the prefactors as matrix elements 
are currently out of reach of our formalism. In the work \cite{FabMc}
scenario (i) was advocated, because it naturally leads to a 
crossover for $\langle\sigma_0^Z \sigma_{n}^Z\rangle$
at a temperature $T_0(n)$ which in dependence on the separation $n$ 
converges to $T_0$ with exponential rate
(formula (2.10) of \cite{FabMc} with three fit parameters). 
For scenario (ii) the dependence
of $T_0(n)$ would be algebraic $T_0+A/n$ with two fit parameters $T_0$
and $A$. In fact, the exponential fit looks more convincing than the
algebraic fit. However, the latter procedure involves one fit parameter
less.

As the implications of the fit procedures are not conclusive
we performed a numerical study of the QTM for finite $N$ 
which already allows
for quite accurate, non-trivial high temperature investigations for values
of $\Delta$ close to 0. Surprisingly, this exercise having been
aimed at discriminating between the two scenarios showed 
for $N=4$ that actually {\it both} take place at
the {\it same} temperature! However, for $N=6$ only scenario (ii) 
is realized. We have performed complete eigenvalue computations 
for much larger numbers $N$ up to 14 and always find scenario (ii) realized.
For these larger numbers of $N$ we have not yet computed the matrix elements
such that we cannot decisively comment on (i) at the moment.

In the following we investigate
scenario (ii) in detail, i.e.
the crossover in the two largest
correlation lengths, or equivalently in the corresponding ``energies''
$\E$ for the lowest state with $\K=0$ and the two lowest states with
$\K=\pm 2\pi/\beta$ which are free and bound states, respectively.
Such a level crossing analysis can of course be performed 
within our approach.

\begin{figure}
\begin{center}
\includegraphics[width=0.5\textwidth,angle=270]{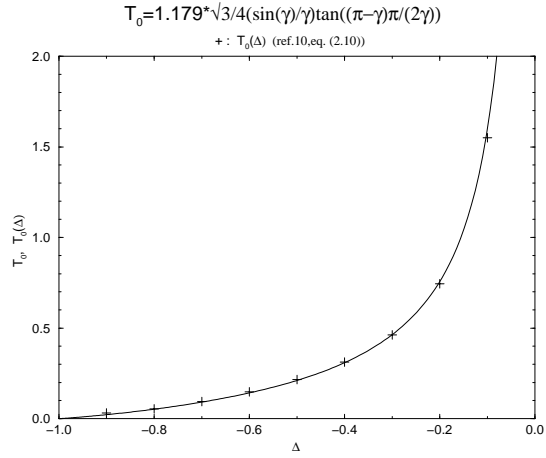}
\end{center}
\caption{Dependence of the crossover temperature $T_0$ as function of
$\Delta$. Crosses denote numerical values, the solid line corresponds 
to the analytic result (\ref{T0}) where a scale factor 1.179 independent
of $\Delta$ has been introduced.}
\label{FigT0}
\end{figure}

The condition for the crossover of a free state
with momenta $\k_1=-\k_2=2\pi/\beta_0$ 
and a bound state with momentum $\kappa=2\pi/\beta_0$ 
where $\beta_0$ corresponds to
crossover temperature $T_0$ is
\bea
\e=\e_1+\e_2 \hbox{ and } 2\k&=&\k_1+\k_2\nonumber\\
\Longrightarrow 2\sinh{\e_i}\sqrt{1-a^2\sinh^2{\e_i}}&=&\sinh\e_i,
\eea
with solution $\sinh^2{\e_i}={3\over4a^2}$. 
Inserting this into $\k_f(\e_i)=2\pi/\beta_0$ yields
\be
T_0={\sqrt{3}\over4}{\sin\gamma\over\gamma}
\tan\left({\pi-\gamma\over\gamma}{\pi\over2}\right).
\label{T0}
\ee
This is in excellent agreement with the numerical analysis of \cite{FabMc},
see Fig.~\ref{FigT0}. An almost perfect agreement with the numerical results
has been achieved by introducing a multiplicative correction close to 1
and independent of $\Delta$. Of course, some correction term had to be
expected as we only calculated within the \lOA.

\begin{figure}
\begin{center}
\includegraphics[width=0.8\textwidth]{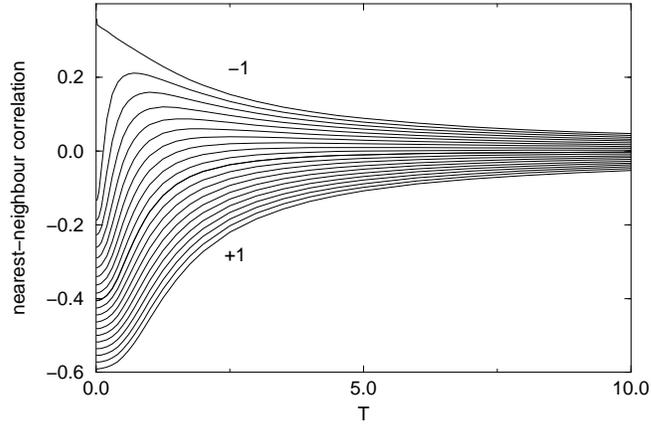}
\end{center}
\caption{Temperature dependence of the nearest-neighbour correlation
$\langle\sigma_j^Z \sigma_{j+1}^Z\rangle$ for 
$\Delta=-1,-0.9,...,0.9,1.0$. Note that for $0\le\Delta\le 1$ and all 
temperatures the values are negative. For $-1<\Delta<0$ there is
a crossover from positive to negative values when passing from high
to low temperatures. For $\Delta=-1$ and all temperatures the
correlation takes positive values which in the limit $T=0$ should
approach the value $+1$ in a singular manner.} 
\label{NNCorr}
\end{figure}

In addition we show ``analytical'' results for the nearest-neighbour
correlations $\langle\sigma_j^Z \sigma_{j+1}^Z\rangle_{T>0}$ derived
as derivatives of the free energy with respect to $\Delta$, see
Fig.~\ref{NNCorr}.

\subsection{Crossover in Macroscopic Quantities}
\begin{figure}
\begin{center}
\includegraphics[width=0.495\textwidth]{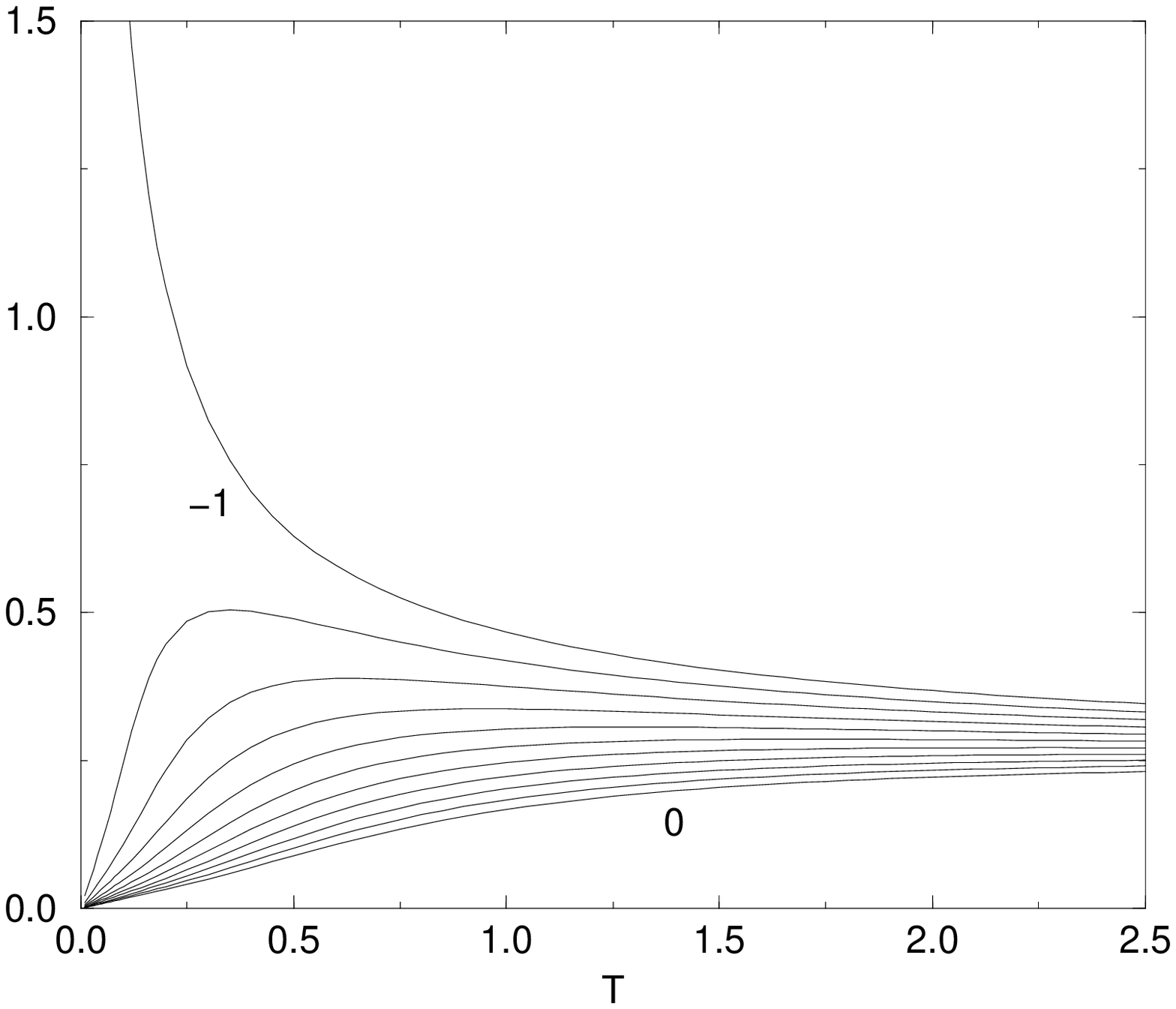}
\includegraphics[width=0.495\textwidth]{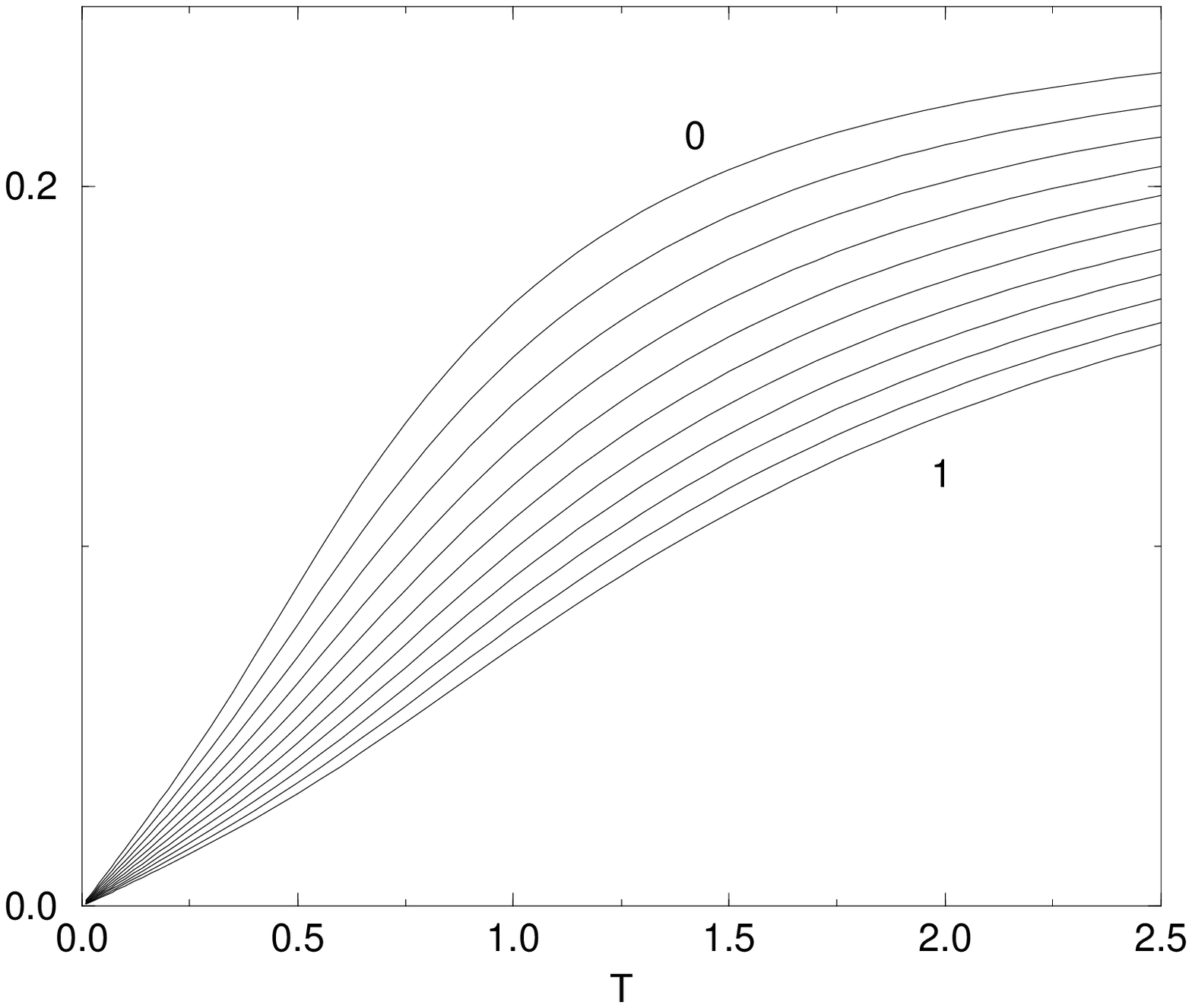}
\end{center}
\caption{Depiction of $\chi(T)\cdot T=\sum_j\langle S_0^Z S_{j+1}^Z\rangle$
for (a) $\Delta=-1, -0.9,...,0$ and (b) $\Delta=0, 0.1,...,1$. Note the
existence of maxima of $\chi(T)\cdot T$ at temperatures of the order $T_0'$
for the attractive case (a). No maxima are observed in the repulsive
case (b).}
\label{FigChiT}
\end{figure}

Finally, we like to present a qualitative argument based on the 
dispersion relations of the elementary excitations of the Hamiltonian
(\ref{spinons},\ref{boundspinons})
to show crossover between ferromagnetic and antiferromagnetic behaviour
at high and low temperatures, respectively. Notice that the
dispersion of free states (\ref{spinons}) is quasislinear, i.e. is 
approximated well by a linear relation as long as the momentum transfer
is less than the reciprocal lattice vector $\pi$. For the bound
states (\ref{boundspinons}) the linear regime is much smaller. For
$0<k<(\pi-\gamma)\pi/2\gamma$ the dispersion
behaves like $\epsilon_b(k)\simeq v\, 2 \sin{k\over 2}$, 
i.e. it is quasi-linear.
For $(\pi-\gamma)\pi/2\gamma<k<\pi$ we find the dependence
$\epsilon_b(k)\simeq v\, 2 a \sin^2{k/ 2}$, 
i.e. quadratic behaviour typical for the
elementary excitations of an isotropic ferromagnetic system. The
crossover takes place for $a\sin{k_c/ 2}=1$ with crossover temperature
\be
T_0'=\epsilon_b(k_c)=2^{3/2}v/a=
{2^{3/2}\pi}{\sin\gamma\over\gamma}
\tan\left({\pi-\gamma\over\gamma}{\pi\over2}\right).
\ee
Note that $T_0'$ is larger than $T_0$ by a factor of 20.52..., though the 
functional dependence on $\Delta$ is identical. We see that both
arguments (crossover of free and bound states of the QTM, and crossover 
from linear to quadratic dispersion of the bound states of the Hamiltonian)
share the same physical origin, but apply to different properties.
In fact, at temperatures
of the order $T_0'$ we see characteristics in the susceptibility data 
which do not exist for the repulsive regime $0<\Delta<1$, see 
Fig.~\ref{FigChiT}. For any $-1<\Delta$ we have
$\lim_{T\to 0}\chi(T)\cdot T=0$, whereas for $\Delta=-1$ the result
is divergent, $\chi(T)\simeq J/(6T^2)$, see \cite{Takahashi,Schlottmann}. 
The noncontinuity of the limiting values is the reason
for the observed temperature maximum
of the quantity $\chi(T)\cdot T$.

For completeness we show results for the specific heat of the $XXZ$ chain
in the attractive as well as repulsive regime.
\begin{figure}
\begin{center}
\includegraphics[width=0.495\textwidth]{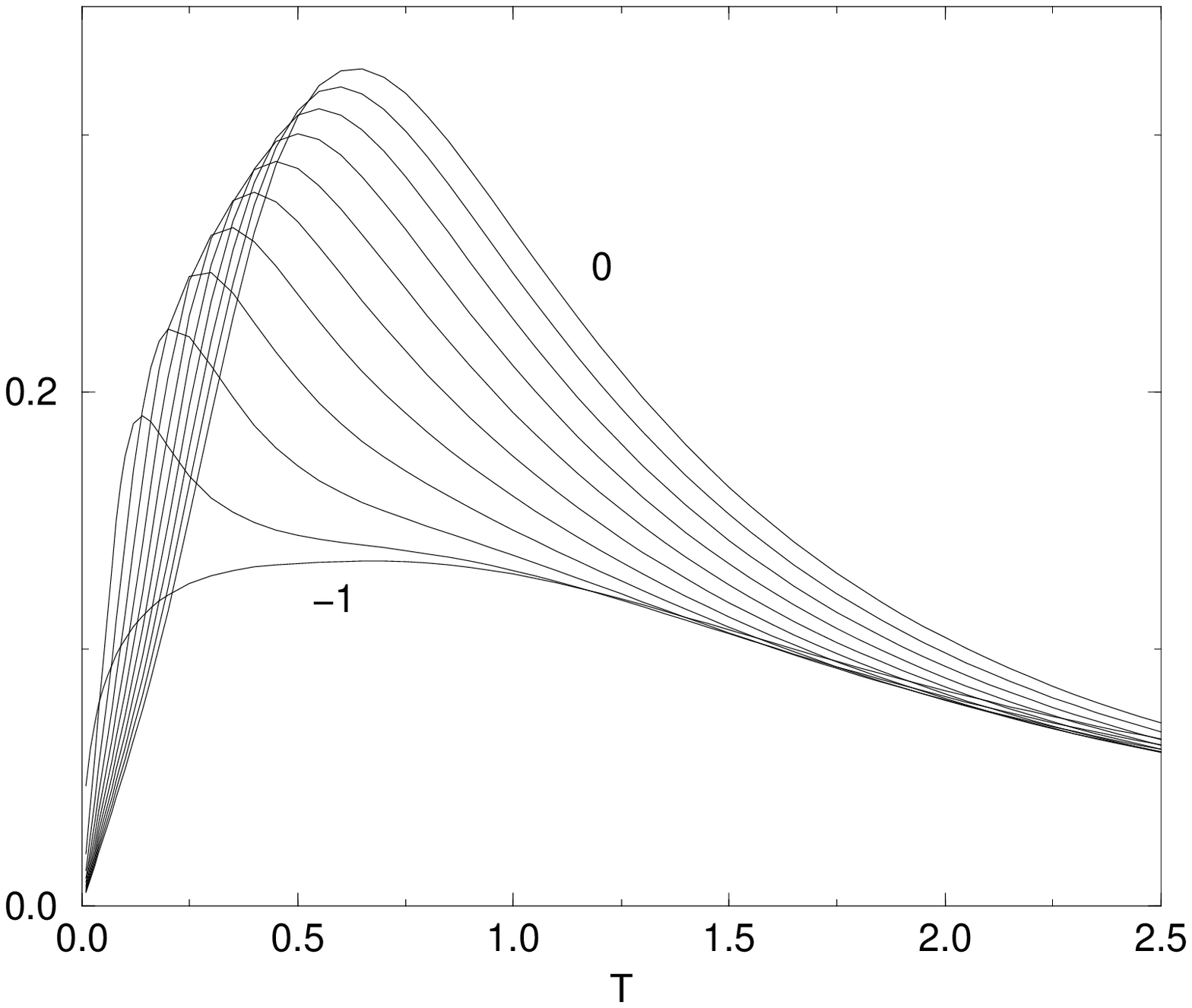}
\includegraphics[width=0.495\textwidth]{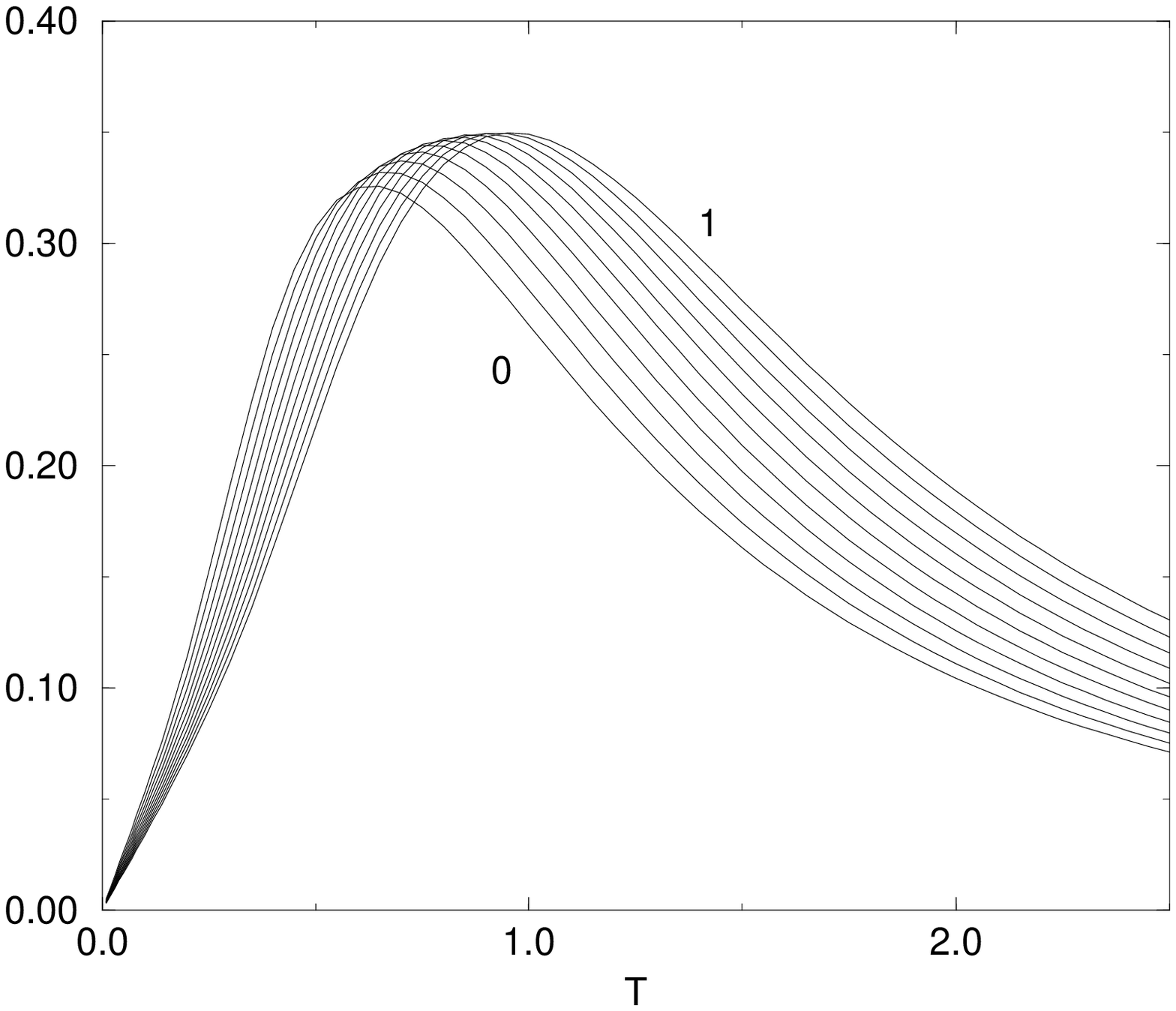}
\end{center}
\caption{Depiction of $c(T)$
for (a) $\Delta=-1, -0.9,...,0$ and (b) $\Delta=0, 0.1,...,1$.}
\label{Figc}
\end{figure}
Note there are no characteristics like finite temperature maxima
in $c(T)$ in addition to the usual ``peak'' typical for spin systems 
with a finite number of degrees per lattice site.

\section{Conclusion}
\label{sec:Concl}
%Add
We have studied the competition of ferromagnetic and 
antiferromagnetic order in the longitudinal correlation
function of the spin-1/2 $XXZ$ chain in the attractive regime.
Within the scenario of level crossing of the quantum transfer matrix
we found an explanation of the sign change in the dominant
exponential asymptotics from positive to negative values when
passing from high to low temperatures. Notably, we found an analytic formula
for the crossover temperature $T_0(\Delta)$ (\ref{T0})
as function of $-1<\Delta<0$
which may be viewed as the maximum temperature below which
conformal field theory is applicable. The agreement of $T_0$ as determined 
numerically and analytically is very good. 

The physical origin  underlying the crossover phenomena in correlation
functions (as well as the magnetic susceptibility) may be viewed in the
existence of bound states. The various manifestations of these peculiar
properties of the spectrum of the quantum transfer matrix (as well as
the Hamiltonian) have been discussed.
A lesson to be drawn from these findings is that the physical properties
of systems possessing bound states are much richer than the usual 
low-energy treatment within bosonization and a subsequent
approximation by Gaussian models.

We want to point out two important questions which remained unanswered.
First, for the explanation of the crossover there are two different scenarios
conceivable. For the numerical analysis a sign change of the coefficient
$A_1$ of the leading term in (\ref{expansion}) was assumed \cite{FabMc},
the analytical work in this paper was based on a level crossing scenario
as also found in explicit numerical treatments of QTMs for
Trotter numbers $4\le N\le 14$. The new level crossing scenario would
suggest a numerical size analysis of the crossover temperature 
somewhat different from that employed in \cite{FabMc}. In preliminary
numerical work we have seen this change to affect the numerical value of
the crossover temperature $T_0$ in only about the second significant 
digit. In our future work we want to focus on a numerical analysis 
of the Hamiltonian \cite{FabMc} and the QTM approach (to be published)
yielding identical numerical values for $T_0$.

Second, even within the level crossing scenario of our analytic
reasoning the result 
(\ref{T0}) represents an analytic approximation to the characteristic 
temperature $T_0$. A comprehensive numerical treatment of 
(\ref{eigexc},\ref{excited}) beyond the (\lOA) approximation
is on the way. 

As a result of the analytical and numerical computations of 
finite $N$--QTM's we expect a decisive answer to the open problems.

\section*{Acknowledgments}
A.K. acknowledges  financial   support  by the   {\it  Deutsche
Forschungsgemeinschaft} under grant  No.   Kl~645/3-1
and support by
the research program of the 
Sonderforschungsbereich 341, K\"oln-Aachen-J\"ulich.
One of us (BMM) is partially supported by the U.S. National Science
Foundation under grant DMR9703543.

%\section*{Appendix}

\end{document}